\def\etal{{\frenchspacing\it et al.}}

\def\beq#1{\begin{equation}\label{#1}}
\def\eeq{\end{equation}}
\def\beqa#1{\begin{eqnarray}\label{#1}}
\def\eeqa{\end{eqnarray}}

\def\bfk{\mbox{\bf k}}
\def\bfr{\mbox{\bf r}}

\def\bfp{\mbox{\bf p}}
\def\bfq{\mbox{\bf q}}

\def\la{\mathrel{\mathpalette\fun <}}

\def\fun#1#2{\lower3.6pt\vbox{\baselineskip0pt\lineskip.9pt
  \ialign{$\mathsurround=0pt#1\hfil##\hfil$\crcr#2\crcr\sim\crcr}}}
  
\documentclass[12pt,preprint]{aastex}
\usepackage{emulateapj5}

\newcommand{\be}{\begin{equation}}
\newcommand{\ee}{\end{equation}}
\newcommand{\ba}{\begin{eqnarray}}
\newcommand{\ea}{\end{eqnarray}}

\shorttitle{Dark Energy and Baryon Acoustic Oscillations}
\shortauthors{Yun Wang}

\begin{document}

\title{Dark Energy Constraints from Baryon Acoustic Oscillations}
\author{Yun~Wang$^{1}$}
\altaffiltext{1}{Department of Physics \& Astronomy, Univ. of Oklahoma,
                 440 W Brooks St., Norman, OK 73019;
                 email: wang@nhn.ou.edu 
                 (Jan 8, 2006)}

\begin{abstract}

Baryon acoustic oscillations (BAO) in the galaxy power spectrum allows us
to extract the scale of the comoving sound horizon at recombination,
a cosmological standard ruler accurately determined by the cosmic
microwave background anisotropy data.
We examine various issues important in the use of BAO to probe dark energy. 
We find that assuming a flat universe, and priors on $\Omega_m$, $\Omega_m h^2$,
and $\Omega_b h^2$ as expected from the Planck mission, the constraints
on dark energy parameters $(w_0,w')$
scale much less steeply with survey area than 
(area)$^{-1/2}$ for a given redshift range.
The constraints on the dark energy density $\rho_X(z)$, however,
do scale roughly with (area)$^{-1/2}$ due to the strong correlation
between $H(z)$ and $\Omega_m$ (which reduces the effect of priors on $\Omega_m$).
Dark energy constraints from BAO are very sensitive to the assumed 
linear scale of matter clustering and the redshift accuracy of the survey. 
For a BAO survey with $0.5\leq z \leq 2$, $\sigma(R)=0.4$ (corresponding to
$k_{max}(z=0)=0.086\,h\,$Mpc$^{-1}$), and $\sigma_z/(1+z)=0.001$,
$({\sigma}_{w_0},{\sigma}_{w'})=$(0.115, 0.183) and (0.069, 0.104) 
for survey areas of 1000 (deg)$^2$ and 10000 (deg)$^2$ respectively.
We find that it is critical to minimize the bias in the
scale estimates in order to derive reliable dark energy constraints.
For a 1000 (10000) square degree BAO survey, a 1$\sigma$ bias in $\ln H(z)$ 
leads to a 2$\sigma$ (3$\sigma$) bias in $w'$. The bias in $w'$ due to 
the same scale bias from $\ln D_A(z)$ is slightly smaller and opposite in sign.
The results from this paper will be useful in assessing different
proposed BAO surveys and guiding the design of optimal dark energy
detection strategies.

\end{abstract}



\section{Introduction}

The most intriguing mystery in cosmology today is the nature of 
the dark energy that is driving the accelerated expansion of the universe
as evidenced by supernova \citep{Riess98,Perl99} and other data.
Although current observational data are consistent with a cosmological
constant (see for example, \cite{WangTegmark04,WangTegmark05,Daly05,Jassal05}),
viable dark energy models abound.
See for example,
\cite{Freese87,Linde87,Peebles88,Wett88,Frieman95,Caldwell98,SH98,Parker99,DGP00,
Mersini01,Freese02,Carroll04,OW04,Alam05,Cai05,Cardone05,Kolb05,MB05}.
\cite{Pad} and \cite{Peebles03} contain reviews of many models.
 
The powerful complementarity of the cosmic microwave anisotropy (CMB)
and galaxy survey data in precision cosmology has long been
noted (see for example, \cite{Bahcall99}, \cite{Eisen99}, and \cite{Wang99}).
An important development in this complementarity is to
use the baryonic acoustic oscillations (BAO) in the galaxy power
spectrum as a cosmological standard ruler to probe dark energy \citep{BG03,SE03},
made possible by accurate determination of the standard ruler scale
by CMB data from WMAP \citep{Bennett03,Spergel03} and 
Planck \footnote{See Planck mission blue book at 
http://www.rssd.esa.int/SA/PLANCK/docs/Bluebook-ESA-SCI(2005)1.pdf}.

The ``wavelength'' of the baryonic acoustic oscillations in $k$-space is determined by 
$k_A=2\pi/s$, where $s$ is the comoving sound horizon at the drag epoch
(when baryons are released from the Compton drag of the 
photons\footnote{For $\Omega_bh^2 \la 0.03$, the drag epoch follows the
last scattering of the photons \citep{EisenHu98}.}),
\be
s=\int_0^{t_{d}} c_s dt/a
=H_0^{-1} \int_{z_{d}}^\infty dz\, \frac{c_s}{
H(z)}
\ee
The sound speed is
\ba
c_s &= &\frac{1}{\sqrt{3(1+R^b)}}, \nonumber\\
R^b &=& \frac{3\rho_b}{4\rho_{\gamma}}=31.5\,\Omega_bh^2 (T_{CMB}/2.7\mbox{K})^{-4}
(z/1000)^{-1},
\ea
and the Hubble parameter is
\be
\label{eq:H(z)}
H(z)=\sqrt{\Omega_m(1+z)^3+ \Omega_{rad}(1+z)^4+ \Omega_k(1+z)^2+
\Omega_X X(z)},
\ee
with $X(z)\equiv \rho_X(z)/\rho_X(0)$ denoting the dark energy density
function. For a cosmological constant, $X(z)=1$ and $\Omega_X =\Omega_\Lambda$.
Since $z_{d}\gg 1$, we have 
\ba
s &\simeq &\frac{H_0^{-1}}{\sqrt{\Omega_m}}
\int_0^{a_{d}} da \,\frac{c_s }{\sqrt{a+a_{eq}}}\nonumber\\
&=&\frac{1}{\sqrt{\Omega_m H_0^2}} \frac{2c}{\sqrt{3z_{eq} R^b_{eq}}}
\, \ln \frac{ \sqrt{1+R^b_{d}}+ \sqrt{R^b_{d}+ R^b_{eq}} }{1+ \sqrt{R^b_{eq}}},
\ea
where ``d'' and ``eq'' refer to the drag epoch and matter-radiation
equality respectively, with
$z_{d}\sim 1089$, and $z_{eq}= 2.5\times 10^4 \Omega_m h^2 
(T_{CMB}/2.7\mbox{K})^{-4}$.
Hence the ``wavelength'' of the baryonic oscillations
depends very strongly on $\Omega_m$, but has negligible dependence
on dark energy (which only became important recently, at redshifts a few
and less). It can therefore be used as a standard ruler to probe dark energy.

The systematic effects of BAO as a standard ruler are: bias between
luminous matter and matter distributions, nonlinear effects, and 
redshift distortions \citep{BG03,SE03}. Cosmological N-body
simulations are required to quantify these effects
\citep{Angulo05,SE05,Springel05,White05}.
Fisher matrix formalism is useful in investigating the impact
of various effects and assumptions that can be parametrized 
in the observed galaxy power spectrum \citep{SE03}.

In this paper, we examine various issues important in the use of 
baryon acoustic oscillations (BAO) to probe dark energy using the
Fisher matrix formalism. 
In particular, we extend the work of Seo \& Eisenstein (2003) by deriving 
the biases in dark energy parameters ($w_0$, $w'$) due to 
systematic errors in the estimated scales.
Throughout this paper, we assume spatial flatness as motivated by inflation
and consistent with current CMB data.

Sec.2 describes the method used in our calculations.
We present results in Sec.3 and summarize in Sec.4.

\section{The method}

\subsection{Error estimation using Fisher matrix}

In estimating the expected errors in dark energy parameters, 
we follow \cite{SE03} in using the Fisher matrix formalism
(see discussion in Sec.4).
We provide details from our calculations for the convenience 
of readers who wish to reproduce our results.

The comoving sizes of an object or feature at redshift $z$ in line-of-sight 
($r_{\parallel}$) and transverse ($r_{\perp}$) directions are
related to the observed sizes $\Delta z$ and $\Delta \theta$ by 
the Hubble parameter $H(z)$ and angular diameter distance $D_A(z)$:
\be
r_{\parallel} = \frac{c\Delta z}{H(z)},
\hskip 1cm
r_{\perp}=(1+z) D_A(z) \Delta \theta,
\ee
The true scale of the baryonic acoustic oscillations (the comoving sound horizon 
at recombination) is known.
Hence if we can measure the ``wavelength'' of the baryonic acoustic oscillations
in the radial direction in successive redshift slices, we obtain
estimates of the cosmic expansion history as a free function of $z$.
While the measurement of the ``wavelength'' of the baryonic acoustic oscillations
in the transverse direction gives us an estimate of the angular diameter 
distance $D_A(z)$ as a free function of $z$.

The accuracy of the power spectrum measurement \citep{Feldman94,Tegmark97} is
\be
\frac{\sigma_P}{P} = 2\pi \sqrt{
\frac{2}{V_{survey} k^2 \Delta k \Delta \mu}} \left( 1+ \frac{1}{nP}\right),
\label{eq:dP/P}
\ee
where $P$ is the average band power, $V_{survey}$ is the total survey volume,
$\Delta k$ is the range of wavenumber $k$ averaged over,
$\Delta \mu$ is the range of the cosine of the angle between the wavevector
{\bf k} and the line of sight,
and $n$ is the comoving number density of observed galaxies.
It is reasonable to assume $nP \sim 3$ \citep{BG03,SE03}.
This gives an estimate of how accurately the ``wavelength'' of the baryonic 
oscillations can be recovered.
Note that in our calculation,
Eq.(\ref{eq:dP/P}) is only used to get an estimate of how accurately
the power spectrum is determined, and {\it not} used to derive
constraints on dark energy and cosmological parameters.

Assuming that the likelihood function for the band powers of a galaxy redshift
survey is Gaussian, the Fisher matrix can be approximated as
\citep{Tegmark97}
\be
F_{ij}= \int_{k_{min}}^{k_{max}}
\frac{\partial\ln P(\bfk)}{\partial p_i}
\frac{\partial\ln P(\bfk)}{\partial p_j}\,
V_{eff}(\bfk)\, \frac{d \bfk^3}{2\, (2\pi)^3}
\label{eq:full Fisher}
\ee
where the derivatives are evaluated at parameter values of the
fiducial model and $V_{eff}$ is the effective volume of the survey
\ba
V_{eff}(k,\mu) &=&\int \left[ \frac{n(\bfr) P(k,\mu)}{ n(\bfr) P(k,\mu)+1}\right]^2
d\bfr^3 \nonumber\\
&=& \left[ \frac{n P(k,\mu)}{ n P(k,\mu)+1}\right]^2 V_{survey},
\ea
where the comoving number density $n$ has been taken to be constant in position.
Here $\mu = \bfk \cdot \hat{\bfr}/k$, with $\hat{\bfr}$ denoting the unit
vector along the line of sight; $\bfk$ is the wavevector with $|\bfk|=k$.
Following \cite{BG03} and \cite{SE03}, we take $k_{min}=0$, and $k_{max}$ given by
requiring that $\sigma(R)\la 0.5$ for $R=\pi/(2k_{max})$ (to ensure that
we are only considering the linear regime).

The observed power spectrum is reconstructed using a particular reference 
cosmology, including the effects of bias and redshift distortions \citep{SE03}:
\ba
\label{eq:P(k)}
P_{obs}(k_{ref\perp},k_{ref\parallel}) &=&
\frac{\left[D_A(z)_{ref}\right]^2  H(z)}{\left[D_A(z)\right]^2 H(z)_{ref}}
\, b^2 \left( 1+\beta\, \frac{ k^2_{\parallel}}{k^2_{\perp}+k^2_{\parallel}}\right)^2
\cdot \nonumber\\
& \cdot& \left[ \frac{G(z)}{G(z=0)}\right]^2 P_{matter}(k|z=0)+ P_{shot}.
\ea
The values in the reference cosmology are denoted by the subscript ``ref'',
while those in the true cosmology have no subscript.
Note that 
\be
k_{ref\perp}=k_\perp D_A(z)/D_A(z)_{ref}, \hskip 0.5cm
k_{ref\parallel}=k_\parallel H(z)_{ref}/H(z).
\label{eq:k}
\ee
The linear redshift distortion $\beta$ is computed from the bias $b$
for fiducial values of the observed galaxy clustering,
$\beta=\Omega_m^{0.6}/b$. $G(z)$ is the linear growth factor.
We normalize the power spectrum to CMB data from COBE \citep{EisenHu98}.

Care needs to be taken in using Eq.(\ref{eq:P(k)}) to
compute the derivatives of $P(\bfk)$
needed for the Fisher matrix in Eq.(\ref{eq:full Fisher}).
Note that $P_{matter}(k|z=0)$ depends on $H(z)$ and $D_A(z)$ through
$k$ (see Eq.(\ref{eq:k})).

For a redshift slice with mean redshift $z$, the estimated parameters (assumed 
to be constant in the redshift slice) are the Hubble parameter $H(z)$, angular 
diameter distance $D_A(z)$, linear redshift distortion $\beta$, linear growth 
function $G(z)$, and an unknown shot noise 
$P_{shot}$.\footnote{This shot noise is the unknown white shot noise that 
remains even after the conventional shot noise of inverse number density has been 
subtracted \citep{SE03}. These could arise from galaxy clustering bias even
on large scales due to local bias \citep{Seljak00}.}
These are estimated simultaneously with  
$\Omega_m$, $\Omega_m h^2$, and $\Omega_b h^2$.
The total number of parameters is $5N+3$ for a BAO survey divided into
$N$ redshift slices.

Note that the same cutoff scales $k_{max}(z)$ need to be used 
for computing {\it all} the Fisher matrix elements, including
those of $\Omega_m$, $\Omega_m h^2$, and $\Omega_b h^2$.
To implement this, calculate $F_{ij}$ in each redshift slice
using the appropriate $k_{max}(z)$,
then sum over all the redshift slices for $\Omega_m$, 
$\Omega_m h^2$, and $\Omega_b h^2$.
If a fixed cutoff scale is used for computing $F_{ij}$
for the cosmological parameters, the degeneracy between
($H(z)$, $D_A(z)$) from the redshift slices and the cosmological
parameters (especially $\Omega_m$) will be artificially broken,
leading to significant under-estimates of errors in
($H(z)$, $D_A(z)$) and the dark energy parameters.

To obtain constraints on ($w_0$, $w'$, $\Omega_m$, $\Omega_m h^2$),
the parameters of interest are $\bfp=(\ln H^i$, $\ln D_A^i$, $i$=1,2, ..., $N$; 
$\Omega_m$, $\Omega_m h^2$).
First marginalize over ($\beta$, $G(z)$, $P_{shot}$) in each redshift slice
as well as $\Omega_b h^2$ by taking the submatrix (of the parameters
of interest) of the inverse of the full Fisher matrix, then
invert it to obtain the Fisher matrix of the parameters of interest,
$F^{sub}$. The Fisher matrix of $\bfq=(w_0$, $w'$, $\Omega_m$, $\Omega_m h^2$)
is obtained by equating the log likelihood functions ($\ln{\cal L}(\bfp)=
\ln{\cal L}(\bfq)$), then taking derivatives with respect to $\bfq$
on both sides. This gives
\be
F_{DE,ij}= \sum_{\alpha,\beta}
\frac{\partial p_\alpha}{\partial q_i}\,
\left( F^{sub}\right)_{\alpha\beta} \,
\frac{\partial p_\beta}{\partial q_j}
\label{eq:F_DE}
\ee

\subsection{Bias in dark energy parameters}

We now extend the work of \cite{SE03} by deriving the bias in dark energy
parameters due to systematic errors 
that bias the extracted standard ruler scale. 

The standard ruler scale is measured using $H(z)$ and $D_A(z)$
in each redshift slice.
First, we marginalize over all other parameters by
taking the ($\ln H^i$, $\ln D_A^i$, $i$=1,2, ..., $N$)
submatrix of the inverse of the full Fisher matrix,
then invert it to obtain the Fisher matrix relevant
for scale determination, $F^{scale}$.

To compute the bias in ($w_0$, $w'$) due to biases in
($\ln H^i$, $\ln D_A^i$, $i$=1,2, ..., $N$),
find the ($w_0$, $w'$) Fisher matrix by 
contracting $F^{scale}$ ($\ln H^i$, $\ln D_A^i$, $i$=1,2, ..., $N$)
to $F_{DE,ij}^{scale}$ ($w_0$, $w'$) (see derivation of Eq.(\ref{eq:F_DE})):
\be
F_{DE,ij}^{scale}= \sum_{\alpha,\beta}
\frac{\partial p_\alpha}{\partial q_i}\,
\left( F^{scale}\right)_{\alpha\beta} \,
\frac{\partial p_\beta}{\partial q_j}
\ee

To relate the biases in ($w_0$, $w'$) to biases in
($\ln H^i$, $\ln D_A^i$, $i$=1,2, ..., $N$),
equate the log likelihood functions 
$\ln{\cal L}^{scale}(\bfp-\bfp_m')=\ln{\cal L}^{scale}(\bfq-\bfq_m')+const.$, subtract 
$\ln{\cal L}^{scale}(\bfp-\bfp_m)=\ln{\cal L}^{scale}(\bfq-\bfq_m)$ from it,
where $\bfp_m'=\bfp_m+\delta\bfp_m$,
and $\bfq_m'=\bfq_m+\delta\bfq_m$ are the biased mean values.
This gives
\ba
(\bfq-\bfq_m)_i^T\,F_{DE,ij}^{scale}\, (\delta\bfq_m)_j &=&
(\bfp-\bfp_m)_\alpha^T\, \left(F^{scale}\right)_{\alpha\beta} \,(\delta\bfp_m)_\beta
\nonumber\\
& &+const.
\ea
Taking derivative with respect to $q_i$ on both sides gives
\be
F_{DE,ij}^{scale}(\delta\bfq_m)_j=\frac{\partial p_\alpha}{\partial q_i}\,
 \left(F^{scale}\right)_{\alpha\beta} \,(\delta\bfp_m)_\beta.
\ee
Hence
\be
(\delta\bfq_m)_i=\left(F_{DE}^{scale}\right)_{ij}^{-1}\,
\frac{\partial p_\alpha}{\partial q_j}\,
 \left(F^{scale}\right)_{\alpha\beta}\, (\delta\bfp_m)_\beta,
 \label{eq:bias w0, w'}
\ee
where $\delta \bfp_m$ are the biases in ($\ln H^i$, $\ln D_A^i$, $i$=1,2, ..., $N$),
and summation is implied over repeated indices.

A known example of BAO systematic bias is 
a bias in the dilation parameter $\alpha=k_{ref}/k_{true}$
for spherically averaged galaxy power spectrum.
\cite{SE05} has shown that $\alpha$ is biased slightly above 1 in
currently used methods for accounting the erasure of baryonic features
due to nonlinear effects (see also \cite{White05}).

\section{Results}

We consider a BAO survey in the redshift range of $0.5\leq z \leq 2$, 
which can be carried out by obtaining the spectra of H$\alpha$
emission line galaxies with spectrographs covering the wavelength
range of 1-2$\,\mu$m.
Note that $0.5\leq z \leq 2$ is the redshift range with the most sensitivity 
to the time variation of dark energy \citep{GB05}, and easily 
accessible by a space mission with simple spectroscopic 
instrumentation. Examples include the BOP MIDEX concept \citep{BOP}, 
and the JEDI mission concept for JDEM \citep{JEDI1,JEDI2}.

Note that our results should qualitatively apply to other BAO surveys,
in particular, ground based surveys.
Ground based BAO surveys, for example, HETDEX \citep{HETDEX} and WFMOS
\citep{WFMOS}, have been planned and will likely occur 
before a BAO survey from space.
Both ground and space BAO surveys will be needed to establish
BAO as a dark energy probe, and to obtain accurate and high precision 
dark energy constraints.

The fiducial cosmological model we have assumed is
$n_S=1$, $h=0.65$, $\Omega_m=0.3$, $\Omega_{\Lambda}=0.7$, $\Omega_b=0.05$.
We assume the following priors as expected from CMB data from the 
Planck mission: $\sigma_{\Omega_m}=0.01$, 
$\sigma_{\Omega_m h^2}/(\Omega_m h^2)=0.01$,
and $\sigma_{\Omega_b h^2}/(\Omega_b h^2)=0.01$.
These parameters are estimated simultaneously with 
$H(z)$, $D_A(z)$, $\beta$, $G(z)$, and $P_{shot}$
from seven redshift slices. Each redshift slice has the thickness
$\Delta z=0.2$, except for the
first redshift slice which has $\Delta z=0.3$ ($0.5<z<0.8$).
We assume a cutoff scale $k_{max}$ given by $k_{max}=\pi/(2R)$,
with $\sigma(R)=0.4$ (except for Fig.2, where $\sigma(R)$ is varied
between 0.3 and 0.5), and a survey redshift accuracy
of $\sigma_z/(1+z)=0.001$ (except for Fig.3, where $\sigma_z/(1+z)$
is varied between 0 and 0.02).
We have conservatively taken the galaxy bias to be 1 (higher bias leads to 
smaller errors in estimated dark energy 
parameters\footnote{This is a small effect when $nP=3$ and
the cosmic variance error dominates.}), and assumed 
the comoving number density of observed galaxies such that
$nP=3$ following \cite{BG03}.

The dark energy equation of state is parametrized by
$w(z)=w_0+w'z$ for $z\leq 2$. It is better to constrain
dark energy density $\rho_X(z)$ instead of dark energy equation of state 
\citep{WangGarnavich01,Tegmark02,WangFreese04}. This is
because dark energy density is more closely related to observables and
thus more tightly constrained. We use dark energy equation of
state parameters in this paper for easy comparison with the work
by others. Note that the dark energy observables here are
$H(z)$, $D_A(z)$ from each redshift slice. The errors in
$H(z)$, $D_A(z)$ are propagated into the errors in 
$w_0$ and $w'$ by contracting the covariance matrix 
(see Eq.(\ref{eq:F_DE})).
We give constraints on $\rho_X(z)$ in Fig.5.

Fig.1 shows the 1$\sigma$ errors in $w_0$ and $w'$ as function of
survey area ($0.5\leq z \leq 2$). In the absence of the priors, the
error bars on dark energy parameters scale as (area)$^{-1/2}$
for a given redshift range. Note that with Planck priors on cosmological
parameters, the constraints on dark energy parameters scale much less 
steeply with survey area than (area)$^{-1/2}$ for a given redshift range.
For a BAO survey with $0.5\leq z \leq 2$, $({\sigma}_{w_0},{\sigma}_{w'})=(0.115,0.183)$ 
and $(0.069,0.104)$ for survey areas of 1000 (deg)$^2$ and 10000 (deg)$^2$ respectively.
The smaller the survey area, the greater the impact of the priors
on cosmological parameters.

Dark energy constraints from BAO are very sensitive to 
the linear scale of matter clustering. 
Fig.2 shows the 1$\sigma$ errors in $w_0$ and $w'$ as functions of
$\sigma(R)$ for survey areas of 1000 and 10000 (deg)$^2$.
The lowest panel in Fig.2 gives the correspondence between $\sigma(R)$
and the cutoff wavenumber $k_{max}$ at $z=0$.
Note that it is important to compare results for the same 
$k_{max}$ at $z=0$, since the same $k_{max}(z=0)$ may correspond to
different $\sigma(R)$ if different normalizations are used for
the power spectrum.

Dark energy constraints from BAO are very sensitive to 
the redshift accuracy of the survey. 
Fig.3 shows the 1$\sigma$ errors in $w_0$ and $w'$ as functions of
$\sigma_z/(1+z)$ for survey areas of 1000 and 10000 square degrees.
The redshift uncertainty $\sigma_z$ leads to a radial smearing
$\sigma_r=c\sigma_z/H(z)$; which modifies
the matter power spectrum in Eq.(\ref{eq:P(k)}) as follows
(see for example \cite{Peacock}):
\be
P_{obs}(\bfk|\sigma_z)=P_{obs}(\bfk|\sigma_z=0)\,
\exp\left(-k^2_{\parallel}\sigma_r^2\right).
\ee

Fig.4 shows $\sigma_H/H$ and $\sigma_{D_A}/D_A$ in each of the
redshift slices, for $\sigma_z/(1+z)=0.001$, 0.01, and 0.02,
for 1000 and 10000 (deg)$^2$ surveys with $0.5\leq z\leq2$.
These represent the accuracy with which model-independent 
constraints on dark energy can be placed.
These are also relevant for estimating likely biases in ($w_0,w'$)
due to biases in scale estimates.

Fig.5 shows the constraints on the dark energy density 
$X(z)=\rho_X(z)/\rho_X(0)$ corresponding to Fig.4 (with the same
assumptions and the same line types).
The errors on $X(z)$ have been propagated from the errors
on $H(z)$ in each redshift slice and the error on $\Omega_m$
(see Eq.[\ref{eq:H(z)}]), with the covariance between $H(z)$
and $\Omega_m$ included.
Note that $X(z)$ does scale roughly as (area)$^{-1/2}$, because 
the strong covariance between $\Omega_m$ and $H(z)$ makes the effects 
of the cosmological priors less important.

Table 1 gives biases in ($w_0,w'$) due to biases in ($\ln H^i$, $\ln D_A^i$)
for 1000 and 10000 square degree BAO surveys with $\sigma(R)=0.4$.
Note that ``1$\sigma$'' denotes setting the bias in ($\ln H^i$, $\ln D_A^i$)
to 1$\sigma$ errors in ($\ln H^i$, $\ln D_A^i$).
The sign of the biases in ($\ln H^i$, $\ln D_A^i$) in Table 1 are chosen
to correspond to scale biases in the same direction, the
direction of wavenumber dilation $\alpha=k_{ref}/k_{true}>1$
as found by \cite{SE05}.

If the dilation is assumed to be uniform in all directions,
then the bias in $\ln H(z)$ and $\ln D_A(z)$ will be equal
and opposite in sign, leading to minimized biases
in $w_0$ and $w'$.
However, the bias in the estimated $\ln H(z)$ should depend on 
the modeling of redshift distortions and on the redshift accuracy
of the survey. Hence it is likely that
dilation in wavenumber will not be uniform in all directions.
For this reason, we give the biases due to $\ln H(z)$ and $\ln D_A(z)$ 
separately in Table 1.
Cosmological N-body simulations are required 
to quantify the expected biases in
$\ln H(z)$ and $\ln D_A(z)$.
 
\section{Discussion and Summary}

Baryon acoustic oscillations (BAO) provide an important new method
for probing dark energy. To extract reliable dark energy constraints
from BAO surveys, we need to understand how the expected errors
depend quantitatively on the assumptions made.

For simplicity and transparency, we use the Fisher information matrix
formalism in this paper. 
Our results are consistent with those of \cite{SE03}, and
agree to about 10\% with those found
using Monte Carlo methods by \cite{BG03} and \cite{GB05},
for the same $k_{max}(z=0)$ and priors on cosmological parameters.
Although Fisher matrix yields the smallest
possible error bars, we have made the most conservative assumptions.
In deriving the dark energy constraints, we marginalize
over many additional physical parameters \citep{SE03}: the linear 
redshift distortion $\beta$, 
linear growth function $G(z)$, and an unknown shot noise $P_{shot}$
in each redshift slice; the matter density ratio $\Omega_m$, 
the matter density $\Omega_m h^2$, and the baryon density $\Omega_b h^2$.
This approach seems to work rather well, as 
\cite{SE05} found that their results from cosmological N-body simulations
are close to what they found using the Fisher matrix formalism \citep{SE03}.

We have examined various issues important in the use of 
BAO to probe dark energy. 
In particular, we extend the work of Seo \& Eisenstein (2003) by deriving 
the biases in dark energy parameters ($w_0$, $w'$) due to 
systematic errors in the estimated scales.

We find that assuming priors on $\Omega_m$, $\Omega_m h^2$,
and $\Omega_b h^2$ as expected from the Planck mission, the constraints
on dark energy parameters $(w_0,w')$ scale much less steeply with survey area (for a
given redshift range) than (area)$^{-1/2}$ (which holds in the absence of the priors).
For a BAO survey with $0.5\leq z \leq 2$, $\sigma(R)=0.4$ (corresponding to
$k_{max}(z=0)=0.086\,h\,$Mpc$^{-1}$), and $\sigma_z/(1+z)=0.001$,
$({\sigma}_{w_0},{\sigma}_{w'})=(0.115,0.183)$ and $(0.069,0.104)$ 
for survey areas of 1000 (deg)$^2$ and 10000 (deg)$^2$ respectively (see Fig.1).
This provides a useful guide in designing optimal survey strategies 
that provide robust and accurate dark energy constraints.

The constraints on the dark energy density $\rho_X(z)$, however,
do scale roughly with (area)$^{-1/2}$ due to the strong correlation
between $H(z)$ and $\Omega_m$ (which reduces the effect of priors on $\Omega_m$).
This is interesting since $\rho_X(z)$ provides direct and
model-independent constraints on dark energy  
\citep{WangGarnavich01,Tegmark02,WangFreese04}.

The BAO constraints on the dark energy equation of state parameters ($w_0$,$w'$)
are very sensitive to the assumed linear scale of matter clustering 
$k_{max}(z)$\footnote{Once realistic N-body simulations are available 
for calibrating the analysis of real data, we will be able to extract
additional information on dark energy parameters by using the data
from the quasi-linear regime.}
and the redshift accuracy of the survey $\sigma_z/(1+z)$ (see Figs.2-3). 
This is important to note since different proposed
BAO surveys are often compared by their expected errors in 
($w_0$,$w'$). Such comparisons are not appropriate unless the same assumptions are 
made about $k_{max}(z)$, and the claimed $\sigma_z/(1+z)$ can be demonstrated to
be feasible. The latter is a key issue for surveys using photometric redshifts
\citep{BG03,SE03,Zhan05}.

It should be noted that the radial BAO are {\it not} observable for
photometric redshift accuracies. The $H(z)$ measurement from using photometric
redshifts is derived from the degree of damping of broad-band 
power in the radial direction on very large scales ($k < 0.05 $h$\,$Mpc$^{-1}$).  
Hence this measurement is much less robust (more susceptible to systematic error) 
than the spectroscopic case, where the radial BAO are detectable
and give $H(z)$ measurements directly. 

We find that it is critical to minimize the bias in the
scale estimates ($\ln H(z)$ and $\ln D_A(z)$) in order to derive 
reliable dark energy constraints.
For a 1000 (10000) sq deg BAO survey,
a 1$\sigma$ bias in $\ln H(z)$ leads to a 2$\sigma$ (3$\sigma$) bias 
in $w'$. The bias in $w'$ due to the same scale bias from $\ln D_A(z)$
is slightly smaller and opposite in sign.
In addition to the scale bias due to the inaccurate modeling of 
nonlinear effects (which impacts the results in the linear regime, 
see \cite{SE05}),
systematic biases in the $H(z)$ measurement could arise when photometric 
redshifts are used.
It is important to quantify the expected bias in
$\ln H(z)$ and $\ln D_A(z)$ using cosmological N-body simulations.
Note that $H(z)$ and $D_A(z)$ are measured as independent parameters.
Since $D_A(z)$ is related to $H(z)$ through an integral for a given
cosmological model in the absence of systematic biases,
comparing the directly measured $D_A(z)$ to the $D_A(z)$ derived
from the $H(z)$ measurement provides us with a cross-check to
help model unknown systematic effects.

The planned BAO surveys from both ground and space telescopes will
play an important role in unraveling the nature of dark energy.
The results from in this paper will be useful in assessing different
proposed BAO surveys and guiding the design of optimal dark energy
detection strategies.

\bigskip

{\bf Acknowledgements}
I am grateful to Chris Blake for making his Monte Carlo results
available to me for comparisons, and for useful discussions.
I thank Max Tegmark, Dan Eisenstein, and Hee-Jong Seo for helpful discussions.
This work was supported in part by NSF CAREER grants AST-0094335.

\begin{table}[htb]
\caption{Biases in ($w_0,w'$) due to biases in ($\ln H^i$, $\ln D_A^i$).}
\begin{center}
\begin{tabular}{cccccc}
\hline\hline
 survey area  & bias in $\ln D_A(z)$ & bias in $\ln H(z)$ & 
 $\sigma_z/(1+z)$  & bias in $w_0$ $(\sigma_{w_0})$&
 bias in $w'$  $(\sigma_{w'})$ \\
\hline
1000 (deg)$^2$ &  -1$\sigma$ & 0. & 0.001 & 0.082 (0.115)   & -0.375 (0.183)\\
		& 	&	& 0.01 & 0.255 (0.266)   & -0.930 (0.727)\\
		& 	&	& 0.02 & 0.273 (0.396)  & -1.021 (1.169)\\
\hline
		 &  0.  & 1$\sigma$ & 0.001  & -0.181 (0.115) & 0.317 (0.183)\\		
		& 	&	& 	0.01 &  -0.303 (0.266)& 0.491 (0.727)\\	
		& 	&	& 0.02  & -0.421 (0.396)&  0.742 (1.169)\\
\hline

10000 (deg)$^2$ &  -1$\sigma$ & 0. & 0.001 & 0.033 (0.069) & -0.291 (0.104)\\
		& 	&	& 0.01 & 0.085 (0.133) & -0.311 (0.251)\\
		& 	&	& 0.02 & 0.089 (0.167) & -0.317 (0.386) \\
\hline
		 &  0.  & 1$\sigma$ & 0.001  & -0.078 (0.069) &  0.263 (0.104)\\		
		& 	&	& 	0.01 & -0.137 (0.133) &  0.172 (0.251)\\	
		& 	&	& 0.02  & -0.157 (0.167) &  0.198 (0.386)\\
\hline		
\end{tabular}
\end{center}
\end{table}

\clearpage
\setcounter{figure}{0}
\plotone{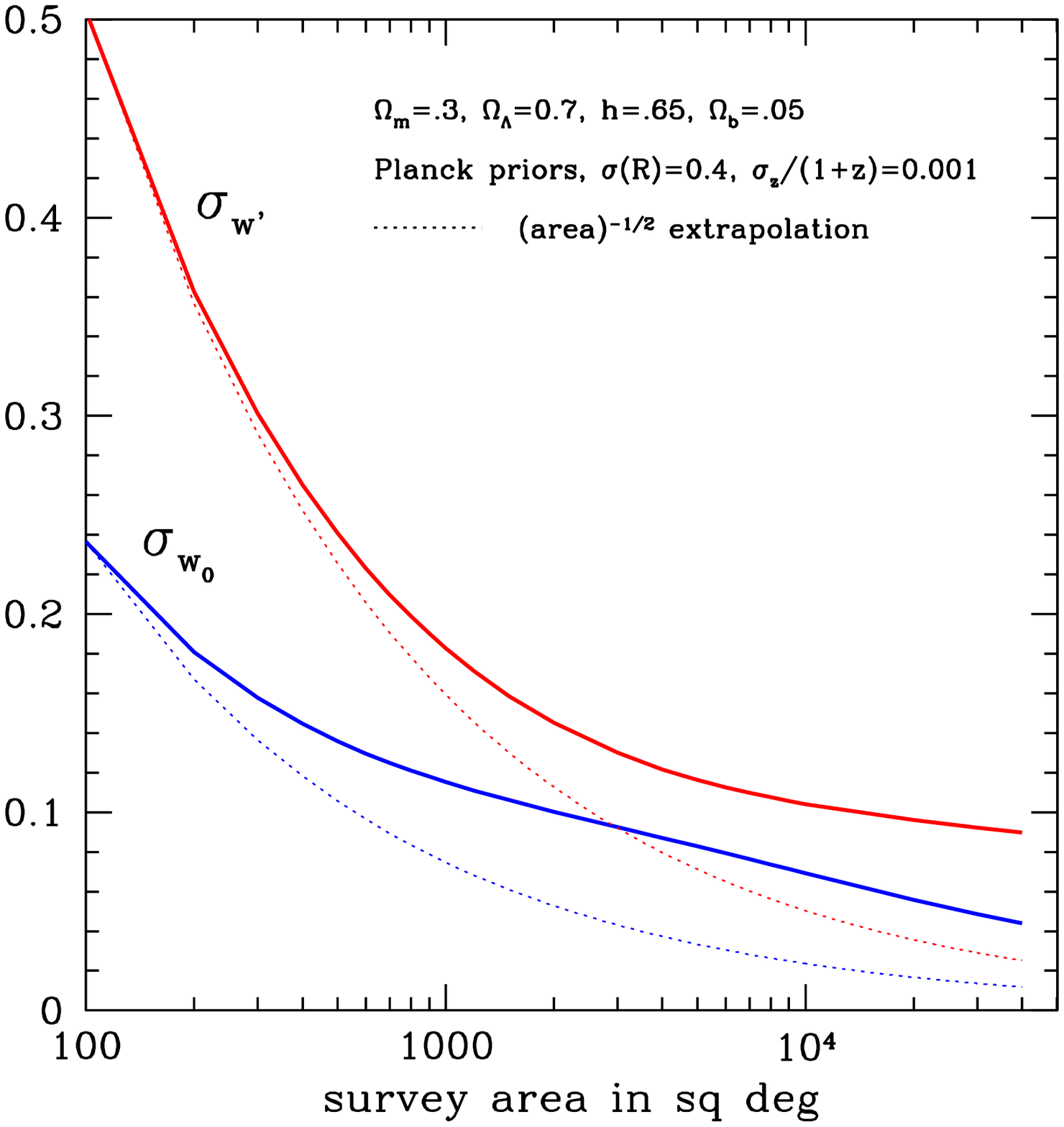}
\figcaption[f1.eps]
{The 1$\sigma$ errors in $w_0$ and $w'$ as function of
survey area ($0.5\leq z \leq 2$). Note that the constraints on dark energy parameters scale much less 
steeply with survey area (for a given redshift range) than (area)$^{-1/2}$ 
(which holds in the absence of the priors).}

\plotone{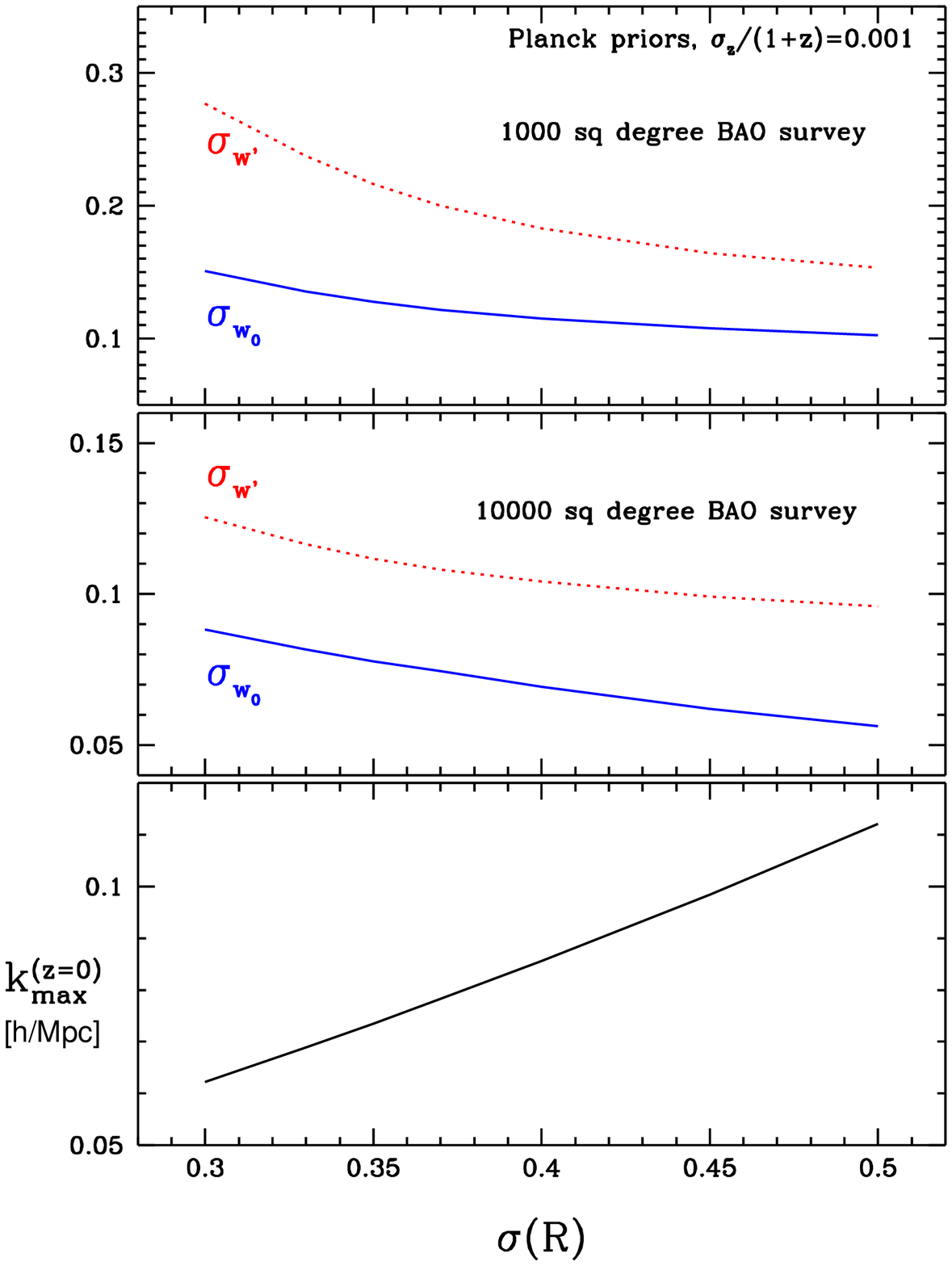}
\figcaption[f2.eps]
{The 1$\sigma$ errors in $w_0$ and $w'$ as function of
$\sigma(R)$ for survey areas of 1000 and 10000 square degrees.
The lowest panel gives the correspondence between $\sigma(R)$
and the cutoff wavenumber $k_{max}$ at $z=0$.}

\plotone{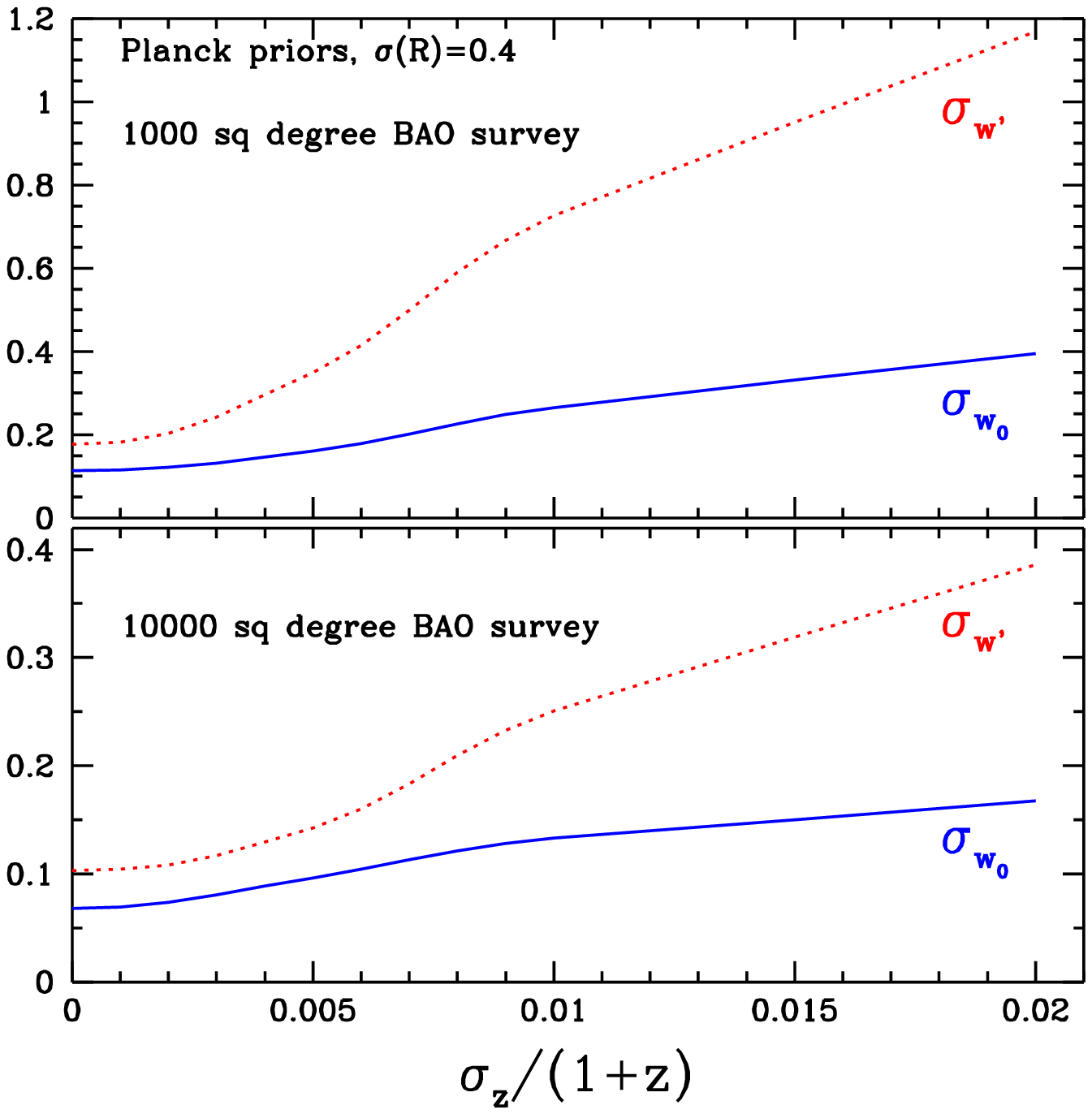}
\figcaption[f3.eps]
{The 1$\sigma$ errors in $w_0$ and $w'$ as function of
$\sigma_z/(1+z)$ for survey areas of 1000 and 10000 square degrees.}

\plotone{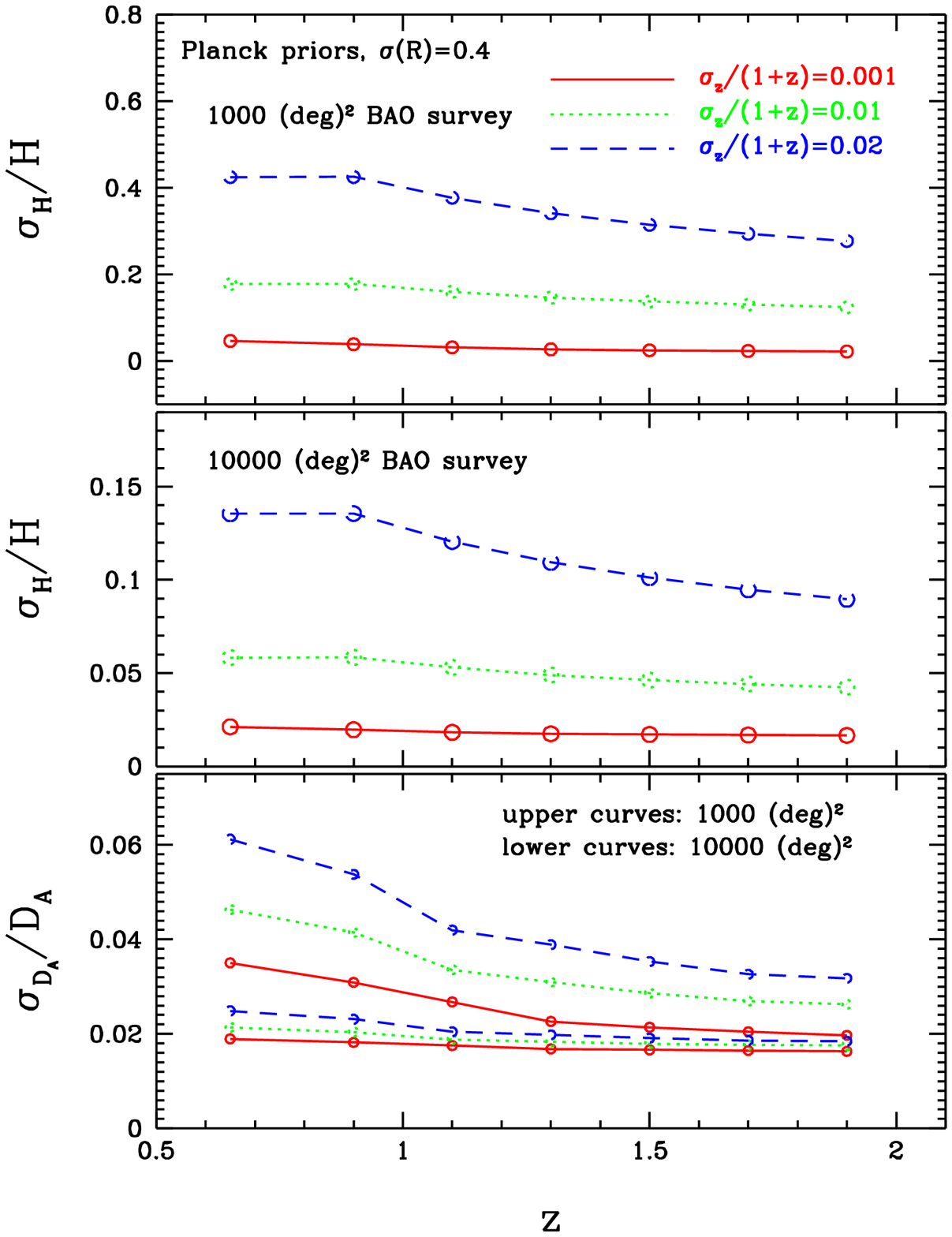}
\figcaption[f4.eps]
{The estimated $\sigma_H/H$ and $\sigma_{D_A}/D_A$ in each of the
redshift slices, for $\sigma_z/(1+z)=0.001$, 0.01, and 0.02
for survey areas of 1000 and 10000 square degrees.
}

\plotone{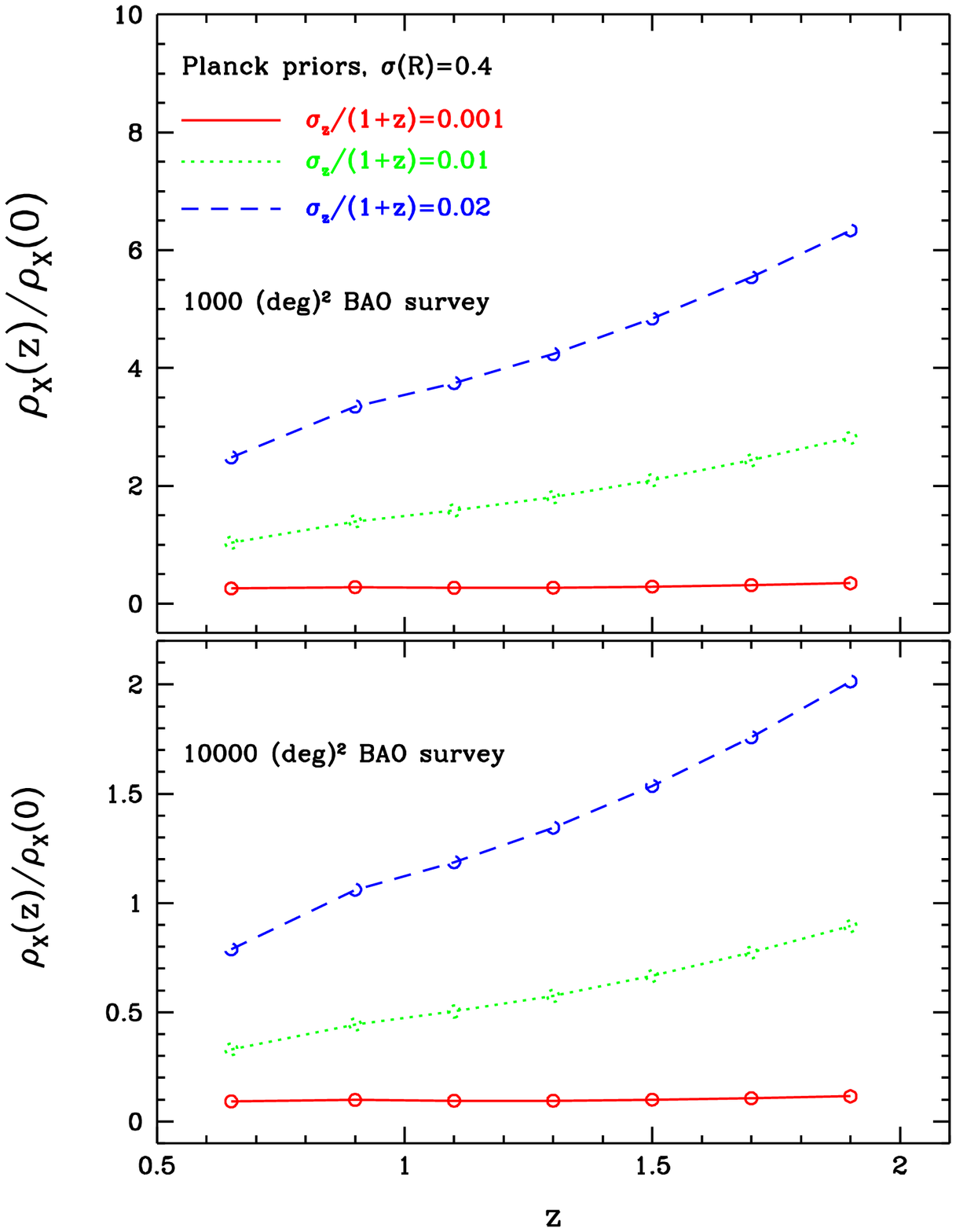}
\figcaption[f5.eps]
{The estimated dark energy density 
$X(z)=\rho_X(z)/\rho_X(0)$ corresponding to Fig.4 (with the same
assumptions and the same line types).}

\end{document}